\documentclass[reprint,twocolumn,aps,prb,superscriptaddress]{revtex4-1}
\usepackage[T1]{fontenc}
\usepackage[utf8]{inputenc}
\usepackage{lmodern}
\usepackage{graphicx}
\usepackage{amsmath}
\usepackage{xcolor}
\usepackage{float}

\begin{document}
\renewcommand{\vec}[1]{\mathbf{#1}}
\newcommand{\ii}{\mathrm{i}}

\title{Spin-orbit-coupled quantum memory of a double quantum dot}

\author{L. \surname{Chotorlishvili}}
\affiliation{Institut f\"ur Physik, Martin-Luther Universit\"at Halle-Wittenberg, D-06120 Halle/Saale, Germany}

\author{A. \surname{Gudyma}}
\affiliation{Max Planck Institute of Microstructure Physics, Weinberg 2, 06120 Halle/Saale, Germany}

\author{J. \surname{W\"atzel}}
\affiliation{Institut f\"ur Physik, Martin-Luther Universit\"at Halle-Wittenberg, D-06120 Halle/Saale, Germany}

\author{A. \surname{Ernst}}
\affiliation{Max Planck Institute of Microstructure Physics, Weinberg 2, 06120 Halle/Saale, Germany}
\affiliation{Institute for Theoretical Physics, Johannes Kepler University, Altenberger Strasse 69, 4040 Linz, Austria}

\author{J. \surname{Berakdar}}
\affiliation{Institut f\"ur Physik, Martin-Luther Universit\"at Halle-Wittenberg, D-06120 Halle/Saale, Germany}

\date{\today}
\begin{abstract}
The concept of quantum memory plays an incisive role in the quantum information theory.
As confirmed by several recent rigorous mathematical studies, the quantum memory inmate in the bipartite system $\rho_{AB}$ can reduce uncertainty about the part $B$, after measurements done on the part $A$.
In the present work, we extend this concept to the systems with a spin-orbit coupling and introduce a notion of spin-orbit quantum memory.
We self-consistently explore Uhlmann fidelity, pre and post measurement entanglement entropy and post measurement conditional quantum entropy of the system with spin-orbit coupling and show that measurement performed on the spin subsystem decreases the uncertainty of the orbital part.
The uncovered effect enhances with the strength of the spin-orbit coupling.
We explored the concept of macroscopic realism introduced by Leggett and Garg and observed that POVM measurements done on the system under the particular protocol are non-noninvasive. For the extended system, we performed the quantum Monte Carlo calculations and explored reshuffling of the electron densities due to the external electric field.
\end{abstract}
\maketitle
\section{Introduction}

Let  us consider a typical setting of a bipartite quantum system~\cite{Alber}, described by the density matrix $\hat{\rho}_{AB}$ shared by two parties Alice ($A$) and Bob ($B$).
Suppose $A$ performs two consecutive measurements of Hermitian observables $X$ and $Y$.
The uncertainty relation in the Robertson's form~\cite{Heisenberg,Robertson} states that the product of the standard deviations is larger or equal to the
expectation value of commutator $\triangle X \triangle Y \geq \frac{1}{2} \big| \langle\psi \big| \big[ X,Y \big] \big| \psi\rangle \big|$ with respect to the shared quantum state $|\psi\rangle$.
From Bob's perspective, the uncertainty in Alice measurement results  depends however on the nature of the quantum state $\big|\psi\rangle$, meaning on whether $\big|\psi\rangle$ is  entangled or separable.
Berta et al.~\cite{Berta} showed that, for Bob, entanglement may  decrease the lower bound for the uncertainty in Alice measurements  outcome.
That means Bob may become more certain about the results of Alice  measurements done on her part $A$, if Bob subsystem $B$ is entangled with $A$.
More specifically, Bob uncertainty concerning Alice measurements is determined by  the quantum conditional entropy defined as follows
$S\big(A|B\big)=S\big(\hat{\varrho}_{AB}\big)-S\big(\rm{tr}_{A}\big(\hat{\varrho}_{AB}\big)\big)$.
Here $\hat{\varrho}_{AB}$ is the post-measurement density matrix of the bipartite system and $S\big(\hat{\varrho}_{AB}\big)$ is the von Neumann entropy $S\big(\hat{\varrho}_{AB}\big)=-{\rm{tr}}\big(\hat{\varrho}_{AB}\ln\big(\hat{\varrho}_{AB}\big)\big)$.
A negative quantum conditional entropy is in contrast to conventional wisdom regarding entropy of a classical system,
as classically, entropy is an extensive quantity and hence the entropy of the whole system should not be lower than the entropy of the subsystem.

Physical realizations of the subsystems $A$ and $B$ are diverse~\cite{Alber}.
For instance, $A$ and $B$ could be  two electrons in a double quantum dot, each hosting spin and orbital degrees of freedom.
The spin and orbital degrees of freedom of electrons may be entangled.
However, due to the SO interaction, spin degrees of freedom may be entangled with orbital degrees as well.
Thus, in a double quantum dot the quantum state of the system may hold spin, orbital and spin-orbit entanglement.
Such solid-state-based systems are very attractive due to their scalability and the various tools at hand to control, read and write information.
When the two dots are in close proximity tunneling sets in, as well as orbital correlation mediated by the Coulomb interaction.
In the presence of a spin orbital (SO) interaction, of the Rashba type~\cite{Rashba} for instance, the spin becomes affected by the orbital motion.
Our interest in this work is devoted to the information obtainable on the orbital subsystem through a measurements done on the spin subsystem and how the quality of this information is affected by the SO interaction.
We note in this context, that the strength of SO interaction in a semiconductor-based quantum structures can be tuned to certain extent by a static electric field.
The orbital part can be assessed for example by exploiting the different relaxation times of the electrons pair to a reservoir depending on their spin state~\cite{Petta,Meunier}.

In what follows, we show that measurement done on the spin subsystem reduces the uncertainty about the orbital part, meaning that information about one subsystem can be extracted indirectly through the measurement done on another subsystem.
We also study the uncertainty of two incompatible measurements done on the spin subsystem and explore factor of quantum memory.
Namely, we prove that when the system is in a pure state, quantum memory reduces the uncertainty of two incompatible spin measurements.

Our focus here is on the  case when Alice does two incompatible quantum measurements on one of the parts of the bipartite system. 
Say, Alice measures two non-commuting spin components of the qubit at her hand. The concept of quantum memory states that the entanglement between qubits of Alice and Bob permits Bob to reduce the upper limit of the uncertainty bound of the measurements done by Alice. 
In what follows we highlight and illustrate by direct numerical examples the subtle effects  of spin-orbital coupling  on quantum memory. 
In particular, spin-orbit-coupled systems may store three different types of entanglements related to spin-spin, spin-orbit, and orbit-orbit parts. 
We prove that only the entire entanglement allows a reduction of the upper bound for the uncertainty. 
After the elimination of the spin-orbit and the orbit-orbit parts, the residual spin-spin entanglement is not enough to reduce the uncertainty. 
Our result is generic and is expected to apply to a broad class of materials with spin-orbital coupling. 

The paper is structured as follows: In section \textbf{\ref{sec:POVM protocols}} we review the experimental studies relevant to our work. Section \textbf{\ref{sec:Model}} presents the theoretical model, in section \textbf{\ref{sec:POVM}} we describe measurement procedures and explore the post-measurement states, in section \textbf{\ref{sec:Uhlmann}} we study the Uhlmann fidelity between pre and post measurement states of the spin subsystem and evaluate post-measurement quantum conditional entropy, in the section \textbf{\ref{sec:entropy}} we study effect of the SO interaction on the quantum discord, in the section \textbf{\ref{sec:witness}} we study non-invasive measurements, in the section \textbf{\ref{sec:Coulomb}} we explore the impact of Coulomb interaction, in the section \textbf{\ref{sec:QMC}} we present results of quantum Monte Carlo calculations for the electron density obtained for the extended system, in the section \textbf{\ref{sec:Memory}} we discuss the problem of quantum memory and conclude the work.

\section{Experimentally feasible POVM protocols in quantum dots}
\label{sec:POVM protocols}

Quantum dots are assumed as an experimental realization to the theory below, similar to the first quantum computing scheme based on spins in isolated quantum dots which was proposed by D. Loss and D. DiVincenzo~\cite{Loss_and_DiVincenzo}, see also~\cite{B.E.Kane,Sukhorukov,HansonBurkard} and references therein.
An ultimate goal of a quantum gate and a quantum information protocol is to read out and record the outcome state.
Several types of local spin measurements were realized experimentally~\cite{Elzerman,2Hanson,HansonAwschalom,Kouwenhoven}. In  quantum dots, the spin can be measured selectively through the spin-to-charge conversion~\cite{Fujisawa, HansonPRL, J.A.Folk, W.LuZ.Ji}.
Our focus is on the experimentally feasible spin POVM (positive operator-valued measure) measurements see \cite{AwschalomLossSamarth}, the only measurement considered throughout the present work.
Fundamental limits for nondestructive measurement of a single spin in a quantum dot was studied recently~\cite{Scalbert}.

Here we briefly look back to experimental and conceptual aspects of the POVM spin measurement in quantum dots.
The spin-resolved filter (barrier) permits to pass through the gate only electrons with particular spin orientation, i.e., transmits $|1\rangle$ and bans the $|0\rangle$.
Thus if particle passes, for sure we know the projection of its spin.
However, what is detected in the experiment is not a spin projection but a charge. Through the change in the electric charge recognized by the electrometer, we infer the information that electron has passed through the filter.
The beauty of this scheme is simpleness that allows introducing POVM projectors  $\Pi_{0}^{A}=|0\rangle\langle0|_{A}$, $\Pi_{1}^{A}=|1\rangle\langle1|_{A}$ for a quantum dot in the formal theoretical discussion.

Of interest is also the single-shot measurement scheme that can selectively access the singlet or the  triplet two-electron  states in a quantum dot~\cite{Meunier}.
The scheme exploits the different coupling strengths of the triplet and singlet states to the reservoir.
Therefore, charge relaxation times are different too $1/\Gamma_{T}<1/\Gamma_{S}$.
A nondestructive measurement is achieved by an electric pulse of duration $\tau$ that shifts temporally the chemical potential of the dot with respect to the Fermi level of the reservoir, where $1/\Gamma_{T}<\tau<1/\Gamma_{S}$ is chosen.
For the dot in the singlet electron state, the time is too short for tunneling, but the triplet state may tunnel.
If two consecutive measurements are done within a time interval shorter than relaxation time $T_{1}$, the measurement procedure is invasive, meaning that the outcome of the second measurement depends on the first measurement.
The measurement procedure is noninvasive if the time interval between measurements exceeds $\Delta\tau_{1,2}>T_{1}$.
In the experiment~\cite{Meunier} values of the parameters for GaAs/Al$_{x}$Ga$_{1-x}$ heterostructure read: $1/\Gamma_{T}=5\mu s$, $\tau=20\mu s$, and $1/\Gamma_{S}=100\mu s$.\\

\section{\label{sec:Model} Model of the system}

The issue of quantum memory has already been addressed for a number of model systems~\cite{Adabi, Cianciaruso, Kumar, Piani, Lecomte, Coles, Winter, Kumar2018}.
Here, we focus particularly on the interacting two-electron double quantum dots
\cite{Efros,rashba2003efficient,Sherman,Sherman2012,Sherman2014,Faniel,Tamborenea,Stepanenko,Pawlowski,Li,hu2000hilbert,Rashba}.
We self-consistently explore the Uhlmann fidelity, pre and post measurement entanglement entropy, and post measurement conditional quantum entropy of the system and show that a measurement performed on the spin subsystem decreases the uncertainty of the orbital part.
This effect becomes more prominent with increasing the strength of SO coupling.

\begin{figure}[t]
\centering
\includegraphics[width=\linewidth]{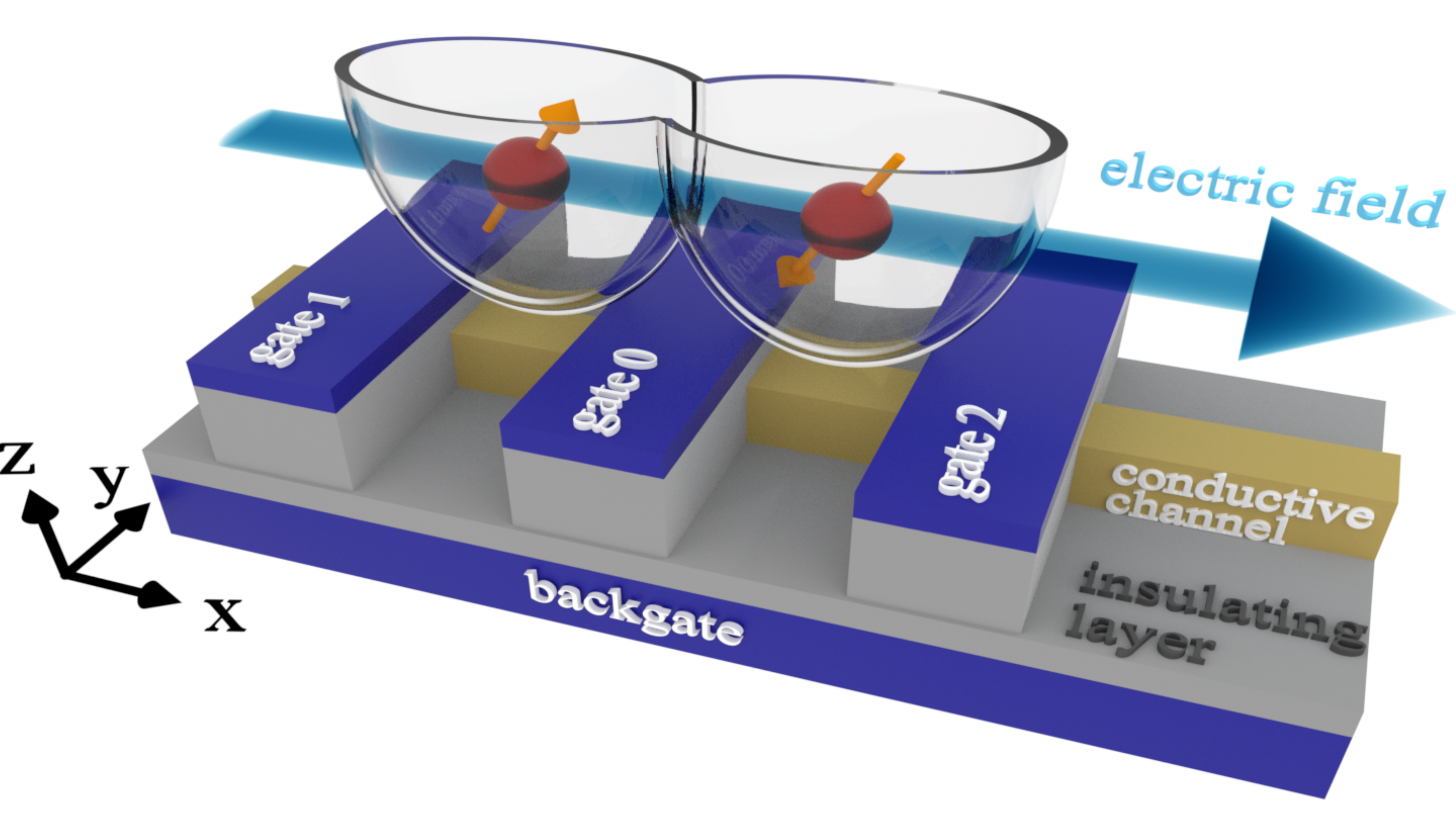}
\caption{Schematic representation of the considered double quantum dot system in the presence of an external electric field and spin-orbit coupling. In a quasi-one-dimensional conductive channel, two quantum dots are created and controlled by two local gates: ”gate 1” and ”gate 2”.
The quantum confinement in $y$-direction is strong enough such that the only lowest subband state $n_y=0$ is occupied. The  ”gate 0” is used to control tunnel junction between the dots (that mimics the changing of the interdot distance). The applied constant electric field, polarized in $x$-direction, is represented by the sky blue arrow.}
\label{fig:scheme}
\end{figure}
We consider a double quantum dot characterized by a rather strong quantum confinement potential in the $y$ and $z$ directions, see pictorial Fig.(\ref{fig:scheme}).
For single particle we use the orbitals $\Psi_{n_x,n_y,n_z}(x,y,z)=N\phi_{n_x}(x)Y_{n_y}(y)Z_{n_z}(z)$
where $\langle\phi_{n_x}|\phi_{n_x'}\rangle=\delta_{n_x,n_x'}$, $\langle Y_{n_y}|Y_{n_y'}\rangle=\delta_{n_y,n_y'}$ and $\langle Z_{n_z}|Z_{n_z'}\rangle=\delta_{n_z,n_z'}$. We consider a situation with a
strong confinement in $y$ and $z$ directions such that only the lowest subbands with $n_y=0$ and $n_z=0$ are occupied.
The relevant dynamics takes place in the $x$ direction only, subject to the effective one-dimensional
potential $V(x)$.
The Hamiltonian of confined electrons reads
\begin{eqnarray}
\label{Hamiltonian0}
&& \hat{H}_{0} =- \frac{\hbar^2}{2m^*}\sum^{N}_{n=1}\frac{\partial^{2}}{\partial x^{2}_{n}}+\sum^{N}_{n<m}V_C(\pmb{r}_n,\pmb{r}_m) +\hat{H}_{\rm SO} \nonumber\\
&& +\sum^{N}_{n=1}\left(V(x_n) + e E x_n\right).
\end{eqnarray}
Here, $V(x)=m^{*}\omega^{2}{\min}\left[(x-\Delta/2)^2,(x+\Delta/2)^2\right]/2$ is the double-dot confinement potential,
\begin{equation}
\hat{H}_{\rm SO}=-i\alpha\sum^{N}_{n=1}\frac{\partial}{\partial{x}_{n}}\hat{\sigma}^y_n + B \sum^{N}_{n=1} \hat{\sigma}^z_n,
\end{equation}
is the Rashba SO term with the magnetic field, $\Delta=\ell d_0$ is the inter-dot distance with the dimensionless scaling factor $\ell$, $m^{*}$ is the electron effective mass, $e$ is the absolute value of the electron charge, and the strength of the constant static electric field is $E$.
The trial magnetic field $B$ applied along the $z$-axis has no particular effect on the phase of the wave function in 1D case but specifies the quantization axis and shifts energy levels.
Note that the Coulomb potential $V_C(\pmb{r}_1,\pmb{r}_2)=e^{2}/{\kappa }\left|\pmb{r}_1-\pmb{r}_2\right|$,
where $\kappa$ is the dielectric constant, still depends on the six coordinates $\pmb{r}_1=(x_1,y_1,z_1)$ and $\pmb{r}_2=(x_2,y_2,z_2)$ and will be reduced to the $(x_{1},x_{2})$-variables later in the text.
In what follows we introduce dimensionless units by setting
$x_{1,2}\rightarrow x_{1,2}/d_{0}$, $\beta=m^*\omega d_{0}^{2}/\hbar$, $\hat{H}_{0}\rightarrow \hat{H}_{0}/(\hbar^{2}/(m^*d_{0}^{2}))$,
$E_0=m^{*}ed_{0}^{3}E/\hbar^{2}$.
We adopt parameters of the semiconductor material GaAs, $\beta=1$ in this case corresponds to a confinement energy
$\hbar\omega=11.4$meV, $d_{0}=10$nm.
Dimensionless electric field $E_{0}=1$ is equivalent to the applied external field of the strength $E=1.1$V/$\mu$m.
The single particle energy levels of a single dot ($\ell=0$) are given then by $\varepsilon_{n}=\beta(n+1/2)$.
For the sake of simplicity to start, we neglect the Coulomb term and treat SO coupling perturbatively.
The antisymmetric total wave-functions are presented as direct products of the orbital and spin parts $\Psi_{n}^{(1)} =\psi^{S}_{n}\otimes\chi_{A}\big(1,2\big)$ and
$\Psi_{n}^{(2)} =\psi^{A}_{n}\otimes\chi_{S}\big(1,2\big)$, where $\big|\chi_{S}^{T_{+}} \big(1,2\big)\big>=|1\uparrow\big>|2\uparrow\big>$,
$\big|\chi_{S}^{T_{-}} \big(1,2\big)\big>=|1\downarrow\big>|2\downarrow\big>$, $|\chi_{S}^{T_{0}} \big(1,2\big)\big>=\frac{1}{\sqrt{2}}\big(\left|1\uparrow\right\rangle\left|2\downarrow\right\rangle+
\left|1\downarrow\right\rangle \left|2\uparrow\right\rangle\big)$ and the asymmetric spin function read $\left|\chi_{A}\big(1,2\big)\right\rangle=
\frac{1}{\sqrt{2}}\big(\left|1\uparrow\right\rangle\left|2\downarrow\right\rangle-\left|1\downarrow\right\rangle\left|2\uparrow\right\rangle\big)$.
We define the two-electron symmetric and antisymmetric coordinate wave functions of a double quantum well as follows:
\begin{equation}
\begin{split}
\label{eigenstates2}
\big|\psi^{S,A}_{n,n'}\big>=\frac{1}{\sqrt{2(1\pm S^2)}} &\left[\psi_{L,n}\big(x_{1}\big)\psi_{R,n'} \big(x_{2}\big)\right.\\
&\left.\pm\psi_{L,n}\big(x_{2}\big)\psi_{R,n'}\big(x_{1}\big)\right],
\end{split}
\end{equation}
where $\psi_{L,n}\big(x\big)$ is the single particle wave function corresponding to the left dot and quantum state $n$, while $\psi_{R,n'}\big(x\big)$ is associated with the right dot and quantum state $n'$ and $S=\langle\psi_{L,n}|\psi_{R,n'}\rangle$ is the overlap integral.
The results of the exact numerical calculations (not shown) have confirmed that for the large values of the parameter $\beta\gg1$ overlap integral is zero $S=0$ and tunneling processes are not activated.
As will be shown bellow effect of the Coulomb term in this case is less relevant and can be neglected safely.
In the double quantum dot, the equilibrium positions of electrons shifts along the $x$-axis by the distance $\pm d_{0}/2$.
The harmonic oscillator eigenfunctions follow Heitler-London ansatz \cite{HeitlerLondon,Fazekas} and read $\psi_{L(R),n}\big(x\big) =\phi_{L(R),n}\big(x\big)$, where
$\phi_{L(R),n}\big(x\big) = \frac{1}{\sqrt{2^{n}n!}}\bigg(\frac{\beta}{\pi}\bigg)^{1/4}\times\exp\left(-\frac{\beta (x\pm 1/2 - d)^{2}}{2}\right)H_{n}\big(x\sqrt{\beta}\big)$, $H_{n}\big(x\sqrt{\beta}\big)$ is Hermite polynomial and $d=eE/d_0m^{*}\omega^2$.

The energy spectrum of unperturbed system is described by the sum of energies of non-interacting oscillators $E_{N}=\beta\big(n+1/2\big)+\beta\big(n'+1/2\big),~~N=(n,n')$ and we introduced the following notations for brevity.
$\big|\Phi_{N}\big>=\big|\psi^{S,A}_{n,n'}\big>\otimes\big|\chi^{A,S}\big>$, see Eq.~\eqref{eigenstates2}.

The presence of the SO term mixes different spin sectors and spin and orbital states.
Considering $\big| \Psi_M \big> = \big|\psi_{0,1}^A\big> \otimes \big|\chi_S^{T_+}\big>$ as an unperturbed wave function we obtain:
\begin{eqnarray}
&&  \big|\Phi_M\big>=\big|\psi_{0,1}^A\big> \otimes \big|\chi_S^{T_+}\big> \nonumber \\
&& + \frac{ \alpha}{2 \sqrt{\beta}} \bigg(\frac{1}{2} \big|\psi^S_{1,1}\big>-\big|\psi^S_{0,0}\big> \bigg)\otimes \big| \chi_A \big>.
\label{eq:total_wf}
\end{eqnarray}
Using Eq.~\eqref{eq:total_wf} and tracing out orbital (spin) parts we construct reduced density matrix of the spin (orbital) subsystem respectively:
$\hat{\rho}^{S}=\quad \frac{1}{Z}\big\{\big|\chi^{T_{+}}_{S}\big>\big<\chi^{T_{+}}_{S}\big|+\frac{5\alpha^{2}}{16\beta}\big|\chi_{A}\big>\big<\chi_{A}\big|\big\}$,
$\hat{\rho}^{or}=\frac{1}{Z}\bigg(\big|\psi_{0,1}^{A}\big>\big<\psi_{0,1}^{A}\big|+\frac{\alpha^{2}}{16\beta}\big|\psi_{1,1}^{S}\big>\big<\psi_{1,1}^{S}\big|+
\frac{\alpha^{2}}{4\beta}\big|\psi_{0,0}^{S}\big>\big<\psi_{0,0}^{S}\big|\bigg)$,
where $Z=1+\frac{5\alpha^2}{16\beta}$.

\section{POVM measurements and post-measurement states}
\label{sec:POVM}

The generic state of two non-interacting particles is a product state.
Therefore a density matrix of a system can be factorized as a
direct product of density matrices of individual particles.
After tracing out states of one particle, product state leaves a system in a pure state with a zero entropy.
However, in the case of fermions, the Pauli principle imposes quantum correlation even in the absence of interaction.
As for an interacting bipartite system in most of the cases, the state is entangled
\cite{suter, Horodecki, Wootters, Olliver, HorodeckiRMP, Zoller, Osterloh, Braunstein}.
Tracing out part of bipartite entangled state results in a mixed state and finite entropy.
Quantum correlation manifests in continuous variables systems as well
\cite{Simon, Weedbrook, Silberhorn, Adesso, Toscano, Serafini, Shchukin, Werner}.
Therefore under certain conditions, we expect the orbital part to be entangled.
After setting theoretical machinery of the problem, we proceed with the
information measures of uncertainty and quantum correlations in the system.
In particular, we specify pre-measurement von Neumann entropy of the orbital and spin subsystems:
$S\big(\hat{\rho}^{or}\big)=-\rm{tr}\big(\hat{\rho}^{or}\ln\big(\hat{\rho}^{or}\big)\big)=-\frac{\alpha^{2}}{16\beta}\ln\big(\frac{\alpha^{2}}{16\beta}\big)-\frac{\alpha^{2}}{4\beta}\ln\big(\frac{\alpha^{2}}{4\beta}\big)$
and $S\big(\hat{\rho}^{s}\big)=-\rm{tr}\big(\hat{\rho}^{s}\ln\big(\hat{\rho}^{s}\big)\big)=-\frac{5\alpha^{2}}{16\beta}\ln\big(\frac{5\alpha^{2}}{16\beta}\big)$ respectively, where we assumed that $Z\approx1$.
Spin and orbital von Neumann entropies increase with the Rashba SO coupling constant $\beta$.
Let us assume that Alice performs POVM measurement \cite{Wilde} on the first qubit at her hand (in what follows we use notations $A=1$ and $B=2$ for the first and second qubit).

After measurement the initial sate collapses either to the post-measurement state
\begin{eqnarray}
\label{post-measurement01}
&& |\Psi_{AB}^{(1)}\big\rangle=\frac{\big(\Pi_{0}^{A}\bigotimes I^{B}\big)\big|\Phi_{M}\big\rangle}{\sqrt{\big\langle \Phi_{M}\big|\big(\Pi_{0}^{A}\bigotimes I^{B}\big)\big|\Phi_{M}\big\rangle}},
\end{eqnarray}
with probability
\begin{eqnarray}
\label{post-measurement02}
&& \Gamma^{A}_{0}=\frac{\langle\Phi_{M}\big|\big(\Pi_{0}^{A}\bigotimes I^{B}\big)\big|\Phi_{M}\big\rangle}{\langle\Phi_{M}\big|\Phi_{M}\big\rangle},
\end{eqnarray}
or to the post-measurement state
\begin{eqnarray}
\label{post-measurement1}
&& |\Psi_{AB}^{(2)}\big\rangle=\frac{\big(\Pi_{1}^{A}\bigotimes I^{B}\big)\big|\Phi_{M}\big\rangle}{\sqrt{\big\langle \Phi_{M}\big|\big(\Pi_{1}^{A}\bigotimes I^{B}\big)\big|\Phi_{M}\big\rangle}},
\end{eqnarray}
with probability
\begin{eqnarray}
\label{post-measurement11}
&& \Gamma^{A}_{1}=\frac{\langle\Phi_{M}\big|\big(\Pi_{1}^{A}\bigotimes I^{B}\big)\big|\Phi_{M}\big\rangle}{\langle\Phi_{M}\big|\Phi_{M}\big\rangle}.
\end{eqnarray}
POVM operators have form: $\Pi_{0}^{A}=|0\rangle\langle0|_{A}$, $\Pi_{1}^{A}=|1\rangle\langle1|_{A}$, $I^{B}$ is the identity operator acting on the qubit $B$.
Easy to see that $\Gamma^{A}_{0}=5\alpha^{2}/\big(10\alpha^{2}+32\beta\big)$; $\Gamma^{A}_{1}=\big(5\alpha^{2}+32\beta\big)/\big(10\alpha^{2}+32\beta\big)$ and $\Gamma^{A}_{1}>\Gamma^{A}_{0}$.
After involved calculations we derive explicit expressions for the post-measurement reduced orbital $\hat{\varrho}_{AB}^{(1,2)}={\rm{tr}}_{s}\big(|\Psi_{AB}^{(1,2)}\big\rangle\big\langle\Psi_{AB}^{(1,2)}|\big)$ and spin $\hat{\sigma}_{AB}^{(1,2)}={\rm{tr}}_{or}\big(|\Psi_{AB}^{(1,2)}\big\rangle\big\langle\Psi_{AB}^{(1,2)}|\big)$ density matrices:

\begin{eqnarray}
\label{post-measurementspin}
&& \hat{\sigma}_{AB}^{(1)}=\big|1\downarrow\rangle|2\uparrow\rangle\langle2\uparrow|\langle1\downarrow|,\nonumber\\
&& \hat{\sigma}_{AB}^{(2)}=\frac{1}{1+5\alpha^{2}/32\beta}\bigg(|1\uparrow\rangle|2\uparrow\rangle\langle2\uparrow|\langle1\uparrow|\nonumber\\
&& +\frac{5\alpha^{2}}{32\beta}|1\uparrow\rangle|2\downarrow\rangle\langle2\downarrow|\langle 1 \uparrow|\bigg),
\end{eqnarray}
and
\begin{eqnarray}
\label{post-measurementorbital}
&& \hat{\varrho}_{AB}^{(1)}=\frac{4}{5}\bigg(\frac{1}{4}|\psi_{1,1}^{s}\rangle\langle\psi_{1,1}^{s}|+|\psi_{0,0}^{s}\rangle\langle\psi_{0,0}^{s}|\nonumber\\
&&- \frac{1}{2}|\psi_{1,1}^{s}\rangle\langle\psi_{0,0}^{s}|-\frac{1}{2}|\psi_{0,0}^{s}\rangle\langle\psi_{1,1}^{s}|\bigg),\nonumber\\
&& \hat{\varrho}_{AB}^{(2)}=\frac{1}{1+5\alpha^{2}/32\beta}\bigg(|\psi_{0,1}^{A}\rangle\langle\psi_{0,1}^{A}|+\frac{\alpha^{2}}{8\beta}\bigg(\frac{1}{4}|\psi_{1,1}^{s}\rangle\langle\psi_{1,1}^{s}|\nonumber\\
&& +|\psi_{0,0}^{s}\rangle\langle\psi_{0,0}^{s}|-\frac{1}{2}|\psi_{1,1}^{s}\rangle\langle\psi_{0,0}^{s}|-\frac{1}{2}|\psi_{0,0}^{s}\rangle\langle\psi_{1,1}^{s}|\bigg)\bigg).
\end{eqnarray}
Since $\alpha/\sqrt{\beta}$ is the small parameter, with high accuracy we set $1+5\alpha^{2}/32\beta\approx1$.

\section{The Uhlmann fidelity and the post-measurement quantum conditional entropy} \label{sec:Uhlmann}

Before study the entropy of the system we explore the fidelity between pre and post measurement states of the spin subsystem.
In its most general form, the fidelity problem was formulated by Uhlmann.
For details about the Uhlmann fidelity, we refer to~\cite{Wilde}.
At first, let us perform the standard purification procedure of the pre $\hat{\rho}^{s}$ and post measurement spin density matrices  $\hat{\sigma}_{AB}$.
We adopt spectral decompositions $\hat{\rho}^{s}=\sum_{x}P_{X}\big(x\big)|x\rangle\langle x|^{AB}$, $\hat{\sigma}_{AB}=\sum_{x}Q_{Y}\big(y\big)|y\rangle\langle y|^{AB}$,
associated with the ensembles $\{P_{X},|x\rangle\}$, $\{Q_{Y},|Y\rangle\}$ where random variables $x,y$ belong to the different alphabets.
A purification with respect to the reference system $R$ we define as follows: $|\phi_{\rho}\rangle^{R,AB}=\sum_{x}\sqrt{P_{X}\big(x\big)}|x\rangle^{R}|x\rangle^{AB}$,
$|\phi_{\sigma}\rangle^{R,B}={\rm{tr}}_{A}\bigg(\sum_{y}\sqrt{Q_{Y}\big(y\big)}|y\rangle^{R}|y\rangle^{AB}\bigg)$.
The Uhlmann fidelity between two mixed states read:
\begin{eqnarray}
\label{Uhlmann fidelity1}
&& F\big({\rm{tr}}_{A}\big(\hat{\sigma}_{AB}\big), \hat{\rho}^{s}\big)= \nonumber \\
&& \textbf{max}_{(U_{\sigma},U_{\rho})}\big|\langle\phi_{\sigma}|\big(U^{\dag}_{\rho}U_{\sigma}\big)^{R}\otimes I^{AB}|\phi_{\sigma}\rangle^{R,AB}|^{2}.
\end{eqnarray}
The Uhlmann theorem \cite{Wilde} facilitates calculation of Uhlmann fidelity and finally, we deduce:
\begin{eqnarray}
\label{Uhlmann fidelity2}
&& F\big({\rm{tr}}_{A}\big(\hat{\sigma}_{AB}^{(2)}\big), \hat{\rho}^{s}\big)=\bigg(1+\frac{5\alpha^{2}}{16\beta\sqrt{2}}\bigg) \nonumber \\
&& \times \frac{1}{1+5\alpha^{2}/32\beta}\times\frac{1}{1+5\alpha^{2}/16\beta}.
\end{eqnarray}
For the small SO coupling we obtain asymptotic estimation:
$F\big({\rm{tr}}_{A}\big(\hat{\sigma}_{AB}^{(2)}\big), \hat{\rho}^{s}\big)\approx1-5\big(3-\sqrt{2}\big)\alpha^{2}/32\beta$.
As we see the distance between pre and post-measurement states decays with SO constant $\alpha$.

\begin{figure*}[!t]
\centering
\includegraphics[width=0.9\linewidth]{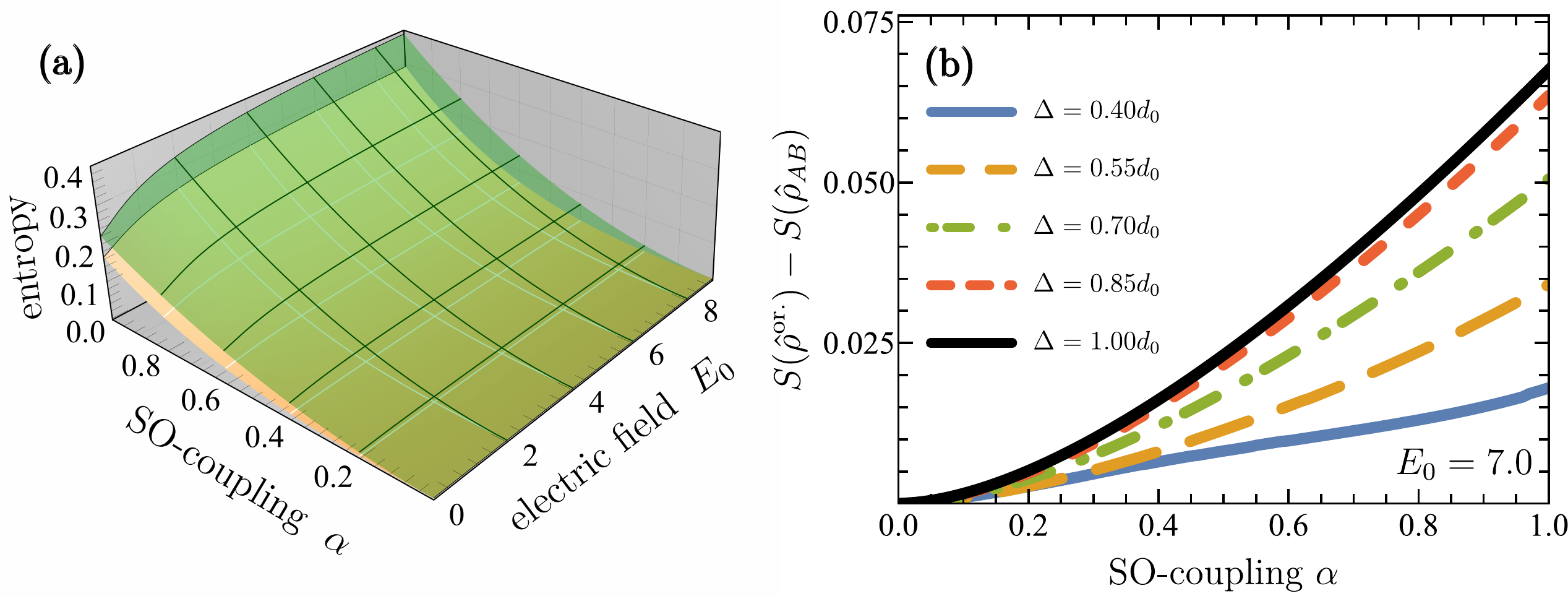}
\caption{Dependence of the von Neumann entropy on the system's and field parameters: (a) Planes describe the pre (green) and post (orange) measurement entropies as a function of the spin-orbit coupling strength $\alpha$ and the applied external electric field $E_{0}$. The effective inter-dot distance is $\Delta=0.8d_0$. (b) The difference between pre and post measurement entropies of the orbital subsystem $S\big(\hat{\rho}^{or}\big)-S\big(\hat{\varrho}_{AB}\big)$ as a function of the spin-orbit coupling $\alpha$, plotted for different inter-dot distances.}
\label{fig:entropy_dist_alpha}
\end{figure*}
Taking into account Eq.~\eqref{post-measurementspin},  Eq.~\eqref{post-measurementorbital} and probabilities Eq.~\eqref{post-measurement02}, Eq.~\eqref{post-measurement11}
we deduce the expression of the post measurement von Neumann entropy of the spin subsystem $S\big(\hat{\sigma}_{AB}^{(2)}\big)=-\Gamma^{A}_{1}\frac{5\alpha^{2}}{32\beta}\ln\big(\frac{5\alpha^{2}}{32\beta}\big)$.
The difference between pre and post measurement entropies of the spin subsystem $S\big(\hat{\sigma}_{S}\big)-S\big(\hat{\sigma}_{AB}\big)=
-\frac{5\alpha^{2}}{16\beta}\ln\big(\frac{5\alpha^{2}}{16\beta}\big)+\Gamma^{A}_{1}\frac{5\alpha^{2}}{32\beta}\ln\big(\frac{5\alpha^{2}}{32\beta}\big)$
is positive for any $\Gamma^{A}_{1}<1$ and that means POVM measurement decreases the entropy of the spin subsystem.
The post measurement von Neumann entropy of the orbital subsystem
$S\big(\hat{\varrho}_{AB}\big)=-\Gamma^{A}_{1}\frac{5\alpha^{2}}{32\beta}\ln\big(\frac{5\alpha^{2}}{32\beta}\big)$.
The difference between pre and post measurement entropies of the orbital subsystem
$S\big(\hat{\rho}^{or}\big)-S\big(\hat{\varrho}_{AB}\big)=-\frac{\alpha^{2}}{16\beta}\ln\big(\frac{\alpha^{2}}{16\beta}\big)-\frac{\alpha^{2}}{4\beta}\ln\big(\frac{\alpha^{2}}{4\beta}\big)+
\Gamma^{A}_{1}\frac{5\alpha^{2}}{32\beta}\ln\big(\frac{5\alpha^{2}}{32\beta}\big)$.
Easy to see that $-\frac{\alpha^{2}}{4\beta}\ln\bigg(\frac{\alpha^{2}}{4\beta}\bigg)>\frac{5\alpha^{2}}{32\beta}\ln\big(\frac{5\alpha^{2}}{32\beta}\big)$
and the entropy after measurement decreases $S\big(\hat{\rho}^{or}\big)-S\big(\hat{\varrho}_{AB}\big)>0$.
An interesting observation is that POVM measurement done on the spin subsystem through the SO interaction decreases the von Neumann entropy of the orbital part.
As larger is SO coupling constant $\alpha$, larger is a decrement of the orbital entropy.
Even more surprising is that measurement equates post-measurement von Neumann entropies of the spin and orbital subsystems $S\big(\hat{\sigma}_{AB}^{(2)}\big)=S\big(\hat{\varrho}_{AB}\big)$.

The pair concurrence of the spin subsystem is defined as follows: $C=\texttt{max}\big(0,\sqrt{R_{1}}-\sqrt{R_{2}}-\sqrt{R_{3}}-\sqrt{R_{4}}\big)$, with
the eigenvalues $R_{n},~n=1,...4$ of the following matrix
$R=\hat{\rho}^{S}\big(\hat{\sigma}_{1}^{y}\otimes \hat{\sigma}_{2}^{y}\big)\big(\hat{\rho}^{S}\big)^{\ast}\big(\hat{\sigma}_{1}^{y}\otimes \hat{\sigma}_{2}^{y}\big)$.
For the pre and post measurement concurrence we obtain: $C\big(\hat{\rho}^{S}\big)=5\alpha^{2}/16\beta$, $C\big(\hat{\sigma}_{AB}\big)=0$.
The measurement disentangles the system.

Taking into account Eq.~\eqref{post-measurementspin}, for the von Neumann entropy of the subsystem $B$ we deduce $S\big({\rm{tr}}_{A}\big(\hat{\sigma}_{AB}^{(2)}\big)\big)=-\Gamma^{A}_{1}\frac{5\alpha^{2}}{32\beta}\ln\big(\frac{5\alpha^{2}}{32\beta}\big)$.
Therefore for the post-measurement conditional quantum entropy we obtain: $S\big(A|B\big)=S\big(\hat{\sigma}_{AB}\big)-S\big({\rm{tr}}_{A}\big(\hat{\sigma}_{AB}\big)\big)=0$.

Note that the conditional quantum entropy of the post-measurement state quantifies the
uncertainty that Bob has about the outcome of Alice's measurement. The zero value of $S\big(A|B\big)$ means that
Bob has precise information about the measurement result. The same effect we see in the post-measurement entropy of the orbital subsystem.
Due to the SO coupling, the measurement done on the qubit $A$ reduces the post-measurement entropy of the orbital subsystem.
The effect of the electric field, Coulomb interaction and tunneling processes activated in case of small inter-dot distance may modify this picture.
In case of a short inter-dot distance (i.e., parameter $\beta$ is an order of $1<\beta<10$) effect of the quantum tunneling processes assisted by the Coulomb interaction becomes important. We explore this problem using numeric methods.

\section{Quantum generalization of conditional entropy}
\label{sec:entropy}

The quantum mutual information quantifies all correlations in the quantum bipartite system, and at least part of these correlations can be classical.
Vedral, Zurek, and others asked the question:
whether it is possible to have a more subtle notion of quantum correlations rather than the entanglement \cite{Vedral, Henderson, Ollivier, Zurek}.
For pure states, the quantum discord is equivalent to the quantum entanglement but is distinct when the state is mixed.
The central issue for the quantum discord is a quantum generalization of conditional entropy (the quantity that is distinct from the conditional quantum entropy).
Quantum discord is quantified as follows:
\begin{widetext}
\begin{eqnarray}
\label{Quantum discord}
D_{A}\big(\hat{\rho}_{s}\big)=\min_{\big\{\Pi_{j}^{B}\big\}}\bigg\{S(A)-S(A,B)-
S\bigg(\sum_{j}p_{j}\rm{tr}\bigg\{\big(\Pi_{j}^{B}\hat{\rho}^{s}\Pi_{j}^{B}\big)/p_{j}\bigg\}\ln\bigg\{\big(\Pi_{j}^{B}\hat{\rho}^{s}\Pi_{j}^{B}\big)/p_{j}\bigg\}\bigg)\bigg\}.
\end{eqnarray}
\end{widetext}
We omit details of calculations and present result for the difference between pre and post-measurement quantum discords
$D_{A}\big(\hat{\rho}^{s}\big)-D_{A}\big(\hat{\sigma}_{AB}\big)=\frac{5\alpha^{2}}{32\beta}\ln4$. From this result we see that
similar to the pre and post-measurement von Neumann entropy, quantum discord decreases after measurement.

\section{Quantum witness and non-invasive measurements}
\label{sec:witness}

The concept of macroscopic realism introduced by Leggett and Garg \cite{LeggettGarg} postulates criteria of noninvasive measurability. In the sequence of two measurements, the first blind measurement has no consequences on the outcome of the second measurement if a system is classical.
However, in the case of quantum systems, any measurement alters the state of the system independently from the fact was the first measurement either blind (i.e., the measurement result is not recorded) or not.
Similar to the Bell's inequalities, quantumness (i.e., entanglement) may violate the macroscopic realism and Leggett-Garg inequalities.
This effect is widely discussed in the literature \cite{SchildClive, WangKne, AvisHayden}.
Quantum witness introduced in \cite{Nori} is the central characteristic of invasive measurements.
In this section we discus particular type of non-invasive measurement protocol.

The directly measured probability we define in terms of the following expression $P_{B}\big(1\big)=\rm{tr}\big\{\Pi^{B}_{1}N\big(\hat{\rho}^{s}\big)\big\}$.
Here $\Pi^{B}_{1}=|1\rangle\langle1|_{B}$ is the operator of the projective measurement done on the second qubit, $N\big(\hat{\rho}^{s}\big)=\sum_{i=1,2}\hat{L}_{i}\hat{\rho}^{s}\hat{L}_{i}^{\dag}$
is the trace preserving quantum channel with Kraus operators $L_{1}=|0\rangle\langle1|_{A}$, $L_{2}=|1\rangle\langle0|_{A}$.
The blind-measurement probability we define as follows: $G_{B}\big(1\big)=\rm{tr}\big\{\Pi^{B}_{1}N\big(\hat{\Xi}^{s}\big)\big\}$, where density matrix of the system after blind measurement is given by
$\hat{\Xi}^{s}=\sum_{i=0,1}\Pi_{i}^{A}\hat{\rho}^{s}\Pi_{i}^{A}$.
The quantum witness that quantifies invasiveness of the quantum measurements is given by the formula:
\begin{equation}
W=|\rm{tr}\big\{\Pi^{B}_{1}\big(N\big(\hat{\rho}^{s}\big)-N\big(\hat{\Xi}^{s}\big)\big)\big\}|.
\label{quantum witness}
\end{equation}
Direct calculations for our system shows that
\begin{equation}
G_{B}\big(1\big)=P_{B}\big(1\big)=\frac{1}{Z}\big(1+5\alpha^{2}/32\beta\big).
\label{quantum witness2}
\end{equation}
The quantum witness is zero $W=0$ indicating that measurements done on the system within this particular procedure are noninvasive.

\section{The effect of the Coulomb interaction}
\label{sec:Coulomb}

We study the case of a short inter-dot distance and the effect of the Coulomb interaction.
We utilize the configurational interaction (CI) ansatz and perform extensive numerical calculations.
Utilizing the single particle orbitals we solve the stationary one-dimensional Schrödinger equation in absence of the Coulomb term.
By means of numerical diagonalization of the single particle Hamiltonian $\hat{H}_{\rm SP}=-\partial_{x}^{2}/2 + V(x) + xE_0$ discretized on a fine space grid we obtain the single-particle orbitals $\phi_i(x)=c_{i,L}\phi_{i,L}(x)+c_{i,R}\phi_{i,R}(x)$ and energies $\varepsilon_i$.
We constructed the symmetric and anti-symmetric two-electron wave functions labeled as $(+,-)$ and evaluate matrix elements of $\hat{H}_0$ including the Coulomb term:
\begin{equation}
\langle\Upsilon_0^{n'}|\hat{H}_0|\Upsilon_0^{n}\rangle = \epsilon^0_n\delta_{n,n'} + \langle\Upsilon_0^{n'}|V_C|\Upsilon_0^{n}\rangle\delta_{b,b'} ,
\label{CI_1}
\end{equation}
where $b$ is a part of the index $n=\left\{i,j,b=(+,-)\right\}$.
Note that two-electron wave-functions $|\Upsilon_0^{n}\rangle$ accounts the effect of doubly-occupied states as well. 
We diagonalize the matrix Eq.~\eqref{CI_1} and obtain the fully correlated two-electron eigenstates and eigenvalues $\left\{|\Psi_n\rangle,\epsilon_n\right\}$.
For a good convergence and reliability of the spectrum, we used 80 single-particle orbitals $|\phi_i\rangle$.
In the last step we add the Rashba SOC term to Eq.~\eqref{Hamiltonian0}.
The matrix elements of the total Hamiltonian including the SO term read
\begin{equation}
\label{CoulombSOtotal}
\begin{split}
\langle\Psi{_{n'\chi'}}|\hat{H}_0+\hat{H}_{\rm SO}|\Psi_{n\chi}\rangle&=\epsilon_n\delta_{n,n'}\delta_{\chi\chi'}\\
&-i\alpha\sum_{i=1}^2\langle\Psi_n'|\partial_{x_i}|\Psi_n\rangle\langle\chi'|\sigma^y_i|\chi\rangle.
\end{split}
\end{equation}
Here the last term corresponds to the Rashba SO interaction in the matrix form. The spin-resolved two-electron eigenstates $|\Phi_{n}\rangle$ and the corresponding energies $\mathcal{E}_n$ we obtain by means of numerical diagonalization of Eq.~\eqref{CoulombSOtotal}. \\

In Fig.~\ref{fig:entropy_dist_alpha} (a) pre and post measurement von Neumann entropies are plotted for the fixed inter-dote distance $\Delta=0.8d_{0}$.
The values of the applied electric field and SO coupling are in the range of $0<E_0<8$ and $0<\alpha<1$, i.e. $E_{0}=1$ corresponds to a static electric field $\approx 1.1 \mbox{ V}/\mu{\rm m}$, $\beta=1$  is equivalent to the realistic parameters adopted for GaAs $\hbar\omega=11,4$meV, $m^*=0,067$m$_{e}$, $d_{0}=10$nm.
The post-measurement von Neumann entropy $S(\hat{\rho}_{AB})$ is always smaller than the pre-measurement entropy $S(\hat{\rho}^{or})$.
Electric field enhances both pre and post-measurement entropies and for $E_{0}>2$ we see the saturation effect.
The difference between pre and post measurement entropies of the orbital subsystem $S(\hat{\rho}^{or})-S(\hat{\rho}_{AB})$ at different inter-dot distances is plotted in Fig.~\ref{fig:entropy_dist_alpha} (b).
As we see measurement done on the spin subsystem reduces the entropy of the orbital part.
Reduction of entropy increases with the strength of SO coupling term $\alpha$.
On the other hand at small inter-dot distances the differences between pre- and post-measurement entropies of the orbital subsystem $S(\hat{\rho}^{or})-S(\hat{\rho}_{AB})$ is smaller due to the Coulomb term.
We note that when SO coupling is zero, the reduced density matrix of the orbital subsystem corresponds to the pure state, and therefore von Neumann entropy is zero see Fig.~\ref{fig:entropy_dist_alpha} (a) and Fig.~\ref{fig:entropy_dist_alpha} (b). The maximum value of the von Neumann entropy depends on the number of the quantum states involved in the process and reaches the peak for the maximally mixed state. Strong electric field increases the amount of the involved quantum states, and von Neumann entropy reaches its saturation value.
Numerical calculations frankly confirm the validity and correctness of analytical results.


\section{Quantum Monte Carlo calculations electron density}
\label{sec:QMC}

Here we consider the extended system \ref{Hamiltonian0} of four electrons in the four-dot confinement potential
\begin{eqnarray}
&&V(x)=\frac{m^{*}\omega^{2}}{2}{\min}\bigg[(x-\frac{3\Delta}{2})^2, (x-\frac{\Delta}{2})^2, (x+\frac{\Delta}{2})^2, \nonumber \\
&&(x+\frac{3\Delta}{2})^2\bigg].
\end{eqnarray}

We perform numerical simulations with the modified continuous spin Variational Monte Carlo (CSVMC) algorithm \cite{SCRVMC1, SCRVMC2}.
We introduce auxiliary spinor vector
\begin{equation}
{\chi}^{\dagger}(s) = \prod_{n=1}^{N} \otimes [e^{is_n}, e^{-is_n}],
\end{equation}
where $s_n$ are auxiliary variables defined on $[0, 2\pi)$ with the periodic boundary conditions.
We construct effective scalar wave-function as a scalar product of the wave-functions and vectors ${\chi}^{\dagger}(s)$ follows
\begin{equation}
    \psi(x,s) = {\chi}^{\dagger}(s) \cdot {\Psi}(x) 
    \label{eq::wf_CSR}
\end{equation}
The inverse transformation is done through the integration over the auxiliary variables
\begin{equation}
    {\Psi}(x) = \frac{1}{(2\pi)^N} \int \prod_{n=1}^{N} d s_n \psi(x,s) {\chi}(s).
\end{equation}
We write the effective Schr\"odinger equation for the scalar wave-function
\begin{equation}
    i \hbar \frac{\partial}{\partial t} \psi(x,s) = \hat{H}_{eff} \psi(x,s),
    \label{eq::Shrodinger_equation_CSR}
\end{equation}
where $\hat{H}_{eff}$ is the effective Hamiltonian.
We construct the effective Hamiltonian replacing the spinor operators by the following operators:
\begin{subequations}
\begin{equation}
    \hat{\sigma}_{x} = \cos(2s) - \sin(2s) \frac{\partial}{\partial s} ,
\end{equation}
\begin{equation}
    \hat{\sigma}_{y} =  \sin(2s) + \cos(2s) \frac{\partial}{\partial s}.
\end{equation}
\begin{equation}
    {\hat{\sigma}}_{z} = - i \frac{\partial}{\partial s},
\end{equation}
\end{subequations}
This transformation expands Hilbert space of the problem from the particular spin sector $s=\frac{1}{2}$ to arbitrary spin.
To select the desired solution from the set of all possible solutions we introduce equality constraints $s^2=\frac{3}{4}$ and $s_z^2=\frac{1}{4}$.
First of these constraints fulfils automatically while second one in introduced directly into the Lagrange function $L = \left\langle \hat{H} \right\rangle + \lambda \left(\left\langle  \hat{\sigma}_z^2 \right\rangle -1\right)$.
The Lagrange function is constructed through minimization of the effective Hamiltonian with the additional spin-variable kinetic energy term.
doing an importance sampling with a guiding wave function $\psi_T$.
We use trial wave-function in the Slater-Jastrow form
\begin{equation}
    \psi_T = D e^J,
\end{equation}
where $J$ is the Jastrow factor which takes into account correlations introduced through the many-body interaction \cite{Jastrow1}.
The none-interacting part is chosen to be a Slater determinant spanned in the lowest lying single-particle orbitals.
Single particle orbitals are approximated with product of Heitler-London functions \cite{HeitlerLondon, Fazekas} and phase calculated from the homogeneous system.

\begin{figure*}[!htb]
\includegraphics[width=0.975\textwidth]{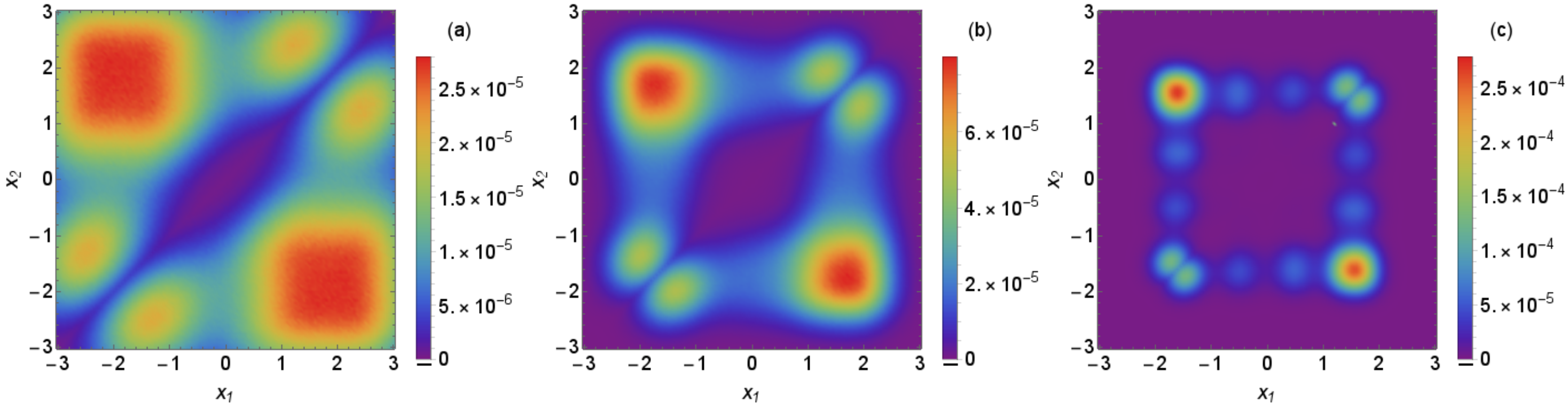}
\caption{The pair distribution function $\rho(x_1,x_2)$ at zero magnetic and electric fields for various values of the trapping parameter $\beta$. The Rashba constant $\alpha=0.4$.
Parameter $\beta$ defines the inverse localization length of wave-function.
When the localization length exceeds the distance between minimums of trapping potential $V(x)$ the electronic wave-function is delocalized.
(3a-b) Delocalized pair distribution function for $\beta = 1$ and $\beta = 3$.
With the increase of $\beta$ potential barrier between minimums of the potential increases and electrons become localized in the minimums of the potential.
(3c) Localized pair distribution function for $\beta =10$.
}
\label{fig:four_well_densities}
\end{figure*}

\begin{figure*}[!htb]
\includegraphics[width=0.975\textwidth]{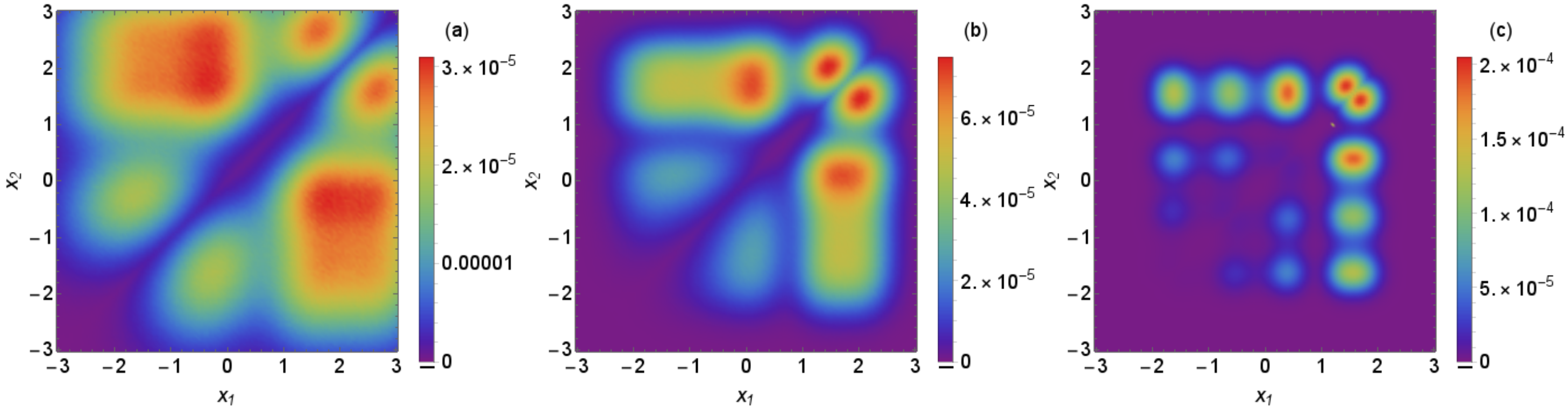}
\caption{ The pair distribution function.
The applied electric field steers the electronic density to the edge of the sample in the direction of the field.
The pair distribution function of the first two particles $\rho(x_1,x_2)$ is centered at the minimums of $V(x)+e E x$ for $E_0 = 1$.  
Various values of the trapping parameter are considered $\beta = 1, 3, 10$.
The Rashba constant is equal to $\alpha=0.4$.
}
\label{fig:four_well_densities_E}
\end{figure*}

In Fig.~\ref{fig:four_well_densities} the pair distribution function is shown for different values of the trapping parameter $\beta = 1, 3$ and $10$.
The Rashba constant is equal to $\alpha=0.4$.
In the regime, $\beta \gg 1$ the electronic density is localized in the vicinity of minimums of the trapping potential and the overlap between neighboring trapping gaps is small (Fig.~\ref{fig:four_well_densities}c).
With the decrease of trapping barrier, electrons delocalize (Figs.~\ref{fig:four_well_densities}a-b).
The effect of the electric field is presented in Fig.~\ref{fig:four_well_densities_E}.
Pair distribution function for $\beta = 1, 3$ and $10$ and $E_0 =1$ is plotted in Fig.~\ref{fig:four_well_densities_E}.
Coordinates $x_1$ and $x_2$ are centered at the minimums of the $V(x)+e E x$. 
At the finite electric field minimums in the direction of the field are energetically preferable and total density shifts towards the direction of the applied field.


\section{Quantum memory}
\label{sec:Memory}

We already showed that measurement done on the spin subsystem reduces the orbital entanglement.
Now we discuss a different scheme when Alice does two incompatible quantum measurements on one of the parts of the bipartite system, and we try to answer the question: whether the spin-orbit interaction can reduce Bob’s total uncertainty about measurements done by Alice?

We consider two cases: in the first case Alice and Bob share the total density matrix of a bipartite SO system Eq.~\eqref{eq:total_wf}
\begin{widetext}
\begin{eqnarray}
\label{total density matrix of a bipartite SO system}
&&\hat{\rho}_{AB}=\frac{1}{Z}\bigg\{\big|\psi_{0,1}^A\big>\big<\psi_{0,1}^A\big|\otimes \big|\chi_S^{T_+}\big>\big<\chi_S^{T_+}\big|+\frac{5\alpha^2}{16\beta}\bigg(\frac{1}{\sqrt{5}}\big|\psi_{1,1}^S\big>-
\frac{2}{\sqrt{5}}\big|\psi_{0,0}^S\big>\bigg)\bigg(\frac{1}{\sqrt{5}}\big<\psi_{1,1}^S\big|-
\frac{2}{\sqrt{5}}\big<\psi_{0,0}^S\big|\bigg)\otimes\big|\chi_A\big>\big<\chi_A\big|\nonumber\\
&&+\frac{\sqrt{5}\alpha}{4\sqrt{\beta}}\bigg(\big|\psi_{0,1}^A\big>\bigg(\frac{1}{\sqrt{5}}\big<\psi_{1,1}^S\big|-
\frac{2}{\sqrt{5}}\big<\psi_{0,0}^S\big|\bigg)\otimes\big|\chi_S^{T_+}\big>\big<\chi_A\big|+\bigg(\frac{1}{\sqrt{5}}\big|\psi_{1,1}^S\big>-
\frac{2}{\sqrt{5}}\big|\psi_{0,0}^S\big>\bigg)\big<\psi_{0,1}^A\big|\otimes\big|\chi_A\big>\big<\chi_S^{T_+}\big|\bigg)\bigg\},
\end{eqnarray}
\end{widetext}
or they share the mixed state formed after tracing the orbital subsystem
$\hat{\rho}^{S}_{AB}=\quad \frac{1}{Z}\big\{\big|\chi^{T_{+}}_{S}\big>\big<\chi^{T_{+}}_{S}\big|+\frac{5\alpha^{2}}{16\beta}\big|\chi_{A}\big>\big<\chi_{A}\big|\big\}$,
where $\big|\chi_{S}^{T_{+}} \big(1,2\big)\big>=|1\uparrow\big>|2\uparrow\big>$, $\big|\chi_{A}\big(1,2\big)\big>=\frac{1}{\sqrt{2}}\big(\left| 1 \uparrow \right\rangle \left| 2 \downarrow \right\rangle-
|1\downarrow\big>|2\uparrow\big>\big)$, $Z=1+\frac{5\alpha^2}{16\beta}$ and the functions $\big|\psi_{0,1}^A\big>$, $\big|\psi_{1,1}^S\big>,~\big|\psi_{0,0}^S\big>$  are defined in the section \textbf{\ref{sec:Model}}.
Bob sends Alice subsystem $A$ and Alice does two incompatible measurements (she measures $\sigma_{A}^{z}$ and $\sigma_{A}^{x}$).
The post-measurement states are given by \cite{Berta}:
\begin{eqnarray}\label{Quantum memory1}
&&\hat{\rho}_{RB}=
\sum\limits_{n}|\psi_{n}\rangle\langle \psi_{n}|\otimes I_{B}  \hat{\rho}^{S}_{AB}
|\psi_{n}\rangle\langle \psi_{n}|\otimes I_{B},
\nonumber\\
&&\hat{\rho}_{QB}=
\sum\limits_{n}|\phi_{n}\rangle\langle \phi_{n}|\otimes I_{B} \hat{\rho}^{S}_{AB}
|\phi_{n}\rangle\langle \phi_{n}|\otimes I_{B}.
\end{eqnarray}
Here $I_{B}$ is the identity operator acting on the subsystem $B$, and $|\psi_{1}\rangle=|1\rangle,~~|\psi_{2}\rangle=|0\rangle$, $|\phi_{1,2}\rangle=\frac{1}{\sqrt{2}}(|0\rangle \pm|1\rangle)$ are the eigenfunctions of $\sigma_{A}^{z}$, $\sigma_{A}^{x}$.
Bob has not precise information about the measurements of Alice.
The uncertainty about outcomes of measurements is quantified through the entropy measure:
\begin{eqnarray}\label{Quantum memory2}
S\big(R|B\big)+S\big(Q|B\big)\geq \ln\bigg(\frac{1}{c}\bigg)+S(A|B\big).
\end{eqnarray}
Here $c=\text{max}_{n,m}|\langle\psi_{m}|\phi_{n}\rangle|^{2}$, $S\big(R|B\big)=-\hat{\rho}_{RB}\ln \hat{\rho}_{RB}+{\rm{tr}}_{R}(\hat{\rho}_{RB})\ln {\rm{tr}}_{R}(\hat{\rho}_{RB})$
is the conditional quantum information, and the last term
$S(A|B\big)$ describes the effect of the quantum memory, meaning that for a negative $S(A|B\big)<0$ quantum memory reduces the uncertainty.
Note that negative conditional quantum entropy points to entanglement in the system.
The inverse statement is not always true, i.e., not for all entangled states, conditional quantum entropy is negative.
Nevertheless, for a pure state Eq.~\eqref{total density matrix of a bipartite SO system} shared by Alice and Bob $\hat{\rho}_{AB}$, the conditional quantum entropy can be calculated explicitly, and it reads:
\begin{eqnarray}\label{Quantum memory pure state}
&&S(A|B\big)_{\hat{\rho}_{AB}}=\frac{5 \alpha^2}{32 \beta Z} \ln\left( \frac{5 \alpha^2}{32 \beta Z}\right)\nonumber\\
&&+\frac{5 \alpha^2 + 32 \beta}{32 \beta Z} \ln\left( \frac{5 \alpha^2 + 32 \beta}{32 \beta Z}\right).
\end{eqnarray}
Easy to see that for any $0<\alpha<\sqrt{\beta}$ conditional quantum entropy is negative for a pure state $S(A|B\big)_{\hat{\rho}_{AB}}<0$.
This fact means that correlations stored in the spin-orbit system work as quantum memory and reduce the uncertainties of measurements.
However in case of the mixed state $\hat{\rho}_{AB}^{S}$ situation is different.
All entropy measures can be calculated analytically, and we deduce:
\begin{widetext}
\begin{eqnarray}
    S(A|B)_{\hat{\rho}_{AB}^{S}} = - \frac{1}{Z } \ln\left( \frac{1}{Z}\right)
    + \frac{5 \alpha^2}{32 \beta Z} \ln\left( \frac{5 \alpha^2}{32 \beta Z}\right)
    - \frac{5 \alpha^2}{16  \beta  Z} \ln\left( \frac{5 \alpha^2}{16 \beta Z}\right)
    + \frac{5 \alpha^2 + 32 \beta}{32 \beta Z} \ln\left( \frac{5 \alpha^2 + 32 \beta}{32 \beta Z}\right),
\end{eqnarray}
\begin{eqnarray}
    S(R|B) = - \frac{1}{Z } \ln\left( \frac{1}{Z}\right)
    - \frac{5 \alpha^2}{32 \beta Z} \ln\left( \frac{5 \alpha^2}{32 \beta Z}\right)
    + \frac{5 \alpha^2 + 32 \beta}{32 \beta Z} \ln\left( \frac{5 \alpha^2 + 32 \beta}{32 \beta Z}\right),
\end{eqnarray}
\begin{eqnarray}
&&S(Q|B) =  \frac{5 \alpha^2}{32 \beta Z} \ln\left( \frac{5 \alpha^2}{32 \beta Z}\right)+\frac{5 \alpha^2 + 32 \beta}{32 \beta Z} \ln\left( \frac{5 \alpha^2 +32 \beta}{32 \beta Z}\right)\nonumber\\
    &&-\frac{5 \alpha^2 + 16 \beta +\sqrt{25 \alpha^4 + 256 \beta^2}}{32 \beta Z} \ln\left( \frac{5 \alpha^2 + 16 \beta +\sqrt{25 \alpha^4 + 256 \beta^2}}{32 \beta Z}\right)\nonumber\\
    &&-\frac{5 \alpha^2 + 16 \beta -\sqrt{25 \alpha^4 + 256 \beta^2}}{32 \beta Z} \ln\left( \frac{5 \alpha^2 + 16 \beta -\sqrt{25 \alpha^4 + 256 \beta^2}}{32 \beta Z}\right).
\end{eqnarray}
\end{widetext}
For strong confinement potential and realistic SO coupling $\alpha/\sqrt{\beta}<1$, $Z=1+\frac{5\alpha^2}{16\beta}\approx1$.
Apparently $S(A|B)_{\hat{\rho}_{AB}^{S}}>0$ meaning that spin orbit coupling in case of a mixed states enhances uncertainties of measurements.
The reason for this nontrivial effect is the following. 
The total entanglement between subsystems $A$ and $B$ stored in the state $\hat{\rho}_{AB}$ consists of spin-spin, spin-orbit, and orbit-orbit contributions. 
Averaging over the orbital states eliminates part of entanglement. 
The residual spin-spin entanglement is not enough to reduce the uncertainty of measurements done by Alice. 
To support this statement, we compare the entanglement stored in the states $\hat{\rho}_{AB}$ and $\hat{\rho}_{AB}^{S}$. 
The reduced density matrix $\hat{\rho}_{A} = \text{tr}_B(\hat{\rho}_{AB})$ has the form: \\
\begin{widetext}
\begin{eqnarray}
&&\hat{\rho}_{A}  = \Big(  \frac{1}{2 Z} \frac{\alpha^2}{32\beta} \big|\psi_{L,1}\big>\big<\psi_{L,1}\big|
    + \frac{1}{2 Z} \frac{\alpha^2}{32\beta} \big|\psi_{R,1}\big>\big<\psi_{R,1}\big|
    + \frac{1}{2 Z} \frac{4 \alpha^2}{32\beta} \big|\psi_{L,0}\big>\big<\psi_{L,0}\big|
    + \frac{1}{2 Z}  \frac{4\alpha^2}{32\beta} \big|\psi_{R,0}\big>\big<\psi_{R,0}\big|
    \Big) \otimes  \big|1 \downarrow\big>\big<1\downarrow\big| \nonumber\\
    &&+
    \left(  \frac{1}{2 Z} \frac{\alpha^2}{32\beta}  \big|\psi_{L,1}\big>\big<\psi_{L,1}\big|
    + \frac{1}{2 Z} \left(1 + \frac{\alpha^2}{32\beta} \right)  \big|\psi_{R,1}\big>\big<\psi_{R,1}\big|
    + \frac{1}{2 Z} \left(1+\frac{4\alpha^2}{32\beta} \right) \big|\psi_{L,0}\big>\big<\psi_{L,0}\big|
    + \frac{1}{2 Z} \frac{4\alpha^2}{32\beta}  \big|\psi_{R,0}\big>\big<\psi_{R,0}\big|
    \right) \nonumber\\
    && \otimes \big|1 \uparrow\big>\big<1\uparrow\big|
\end{eqnarray}
\end{widetext}

The corresponding von Neumann entropy:
\begin{widetext}
\begin{eqnarray}
    && S(\hat{\rho}_{A} ) = - \frac{3}{2Z} \frac{\alpha^2}{32\beta}\ln \left(\frac{1}{2 Z} \frac{\alpha^2}{32\beta}\right)
    - \frac{3}{2Z} \frac{4\alpha^2}{32\beta} \ln\left(\frac{1}{2 Z} \frac{4\alpha^2}{32\beta}\right)
    - \frac{1}{2 Z} \left(1 + \frac{\alpha^2}{32\beta} \right)  \ln\left(\frac{1}{2 Z} \left(1 + \frac{\alpha^2}{32\beta} \right) \right)
    \nonumber\\
    && - \frac{1}{2 Z} \left(1+\frac{4\alpha^2}{32\beta} \right) \ln\left(\frac{1}{2 Z} \left(1+\frac{4\alpha^2}{32\beta} \right)\right).
\end{eqnarray}
\end{widetext}
The von Neumann entropy for the state $\hat{\rho}_{A}^{s} = \text{tr}_B(\hat{\rho}_{AB}^{s})$:
\begin{widetext}
\begin{equation}
    S(\hat{\rho}_A^S) = - \frac{1}{Z} \left(1 + \frac{5\alpha^{2}}{32\beta} \right) \ln\left(\frac{1}{Z} \left(1 + \frac{5\alpha^{2}}{32\beta} \right)\right) - \frac{5\alpha^{2}}{32\beta Z} \ln\left(\frac{5\alpha^{2}}{32\beta Z}\right).
\end{equation}
\end{widetext}
Apparently $S(\hat{\rho}_{A} )>S(\hat{\rho}_{A}^{s} )$ and part of entanglement is lost after averaging over the orbital states.

\section{Conclusions}
Combining the analytical method with extensive numeric calculations, in the present work, we studied the influence of the spin-orbit interaction on the effect of quantum memory.
We observed that measurement done on the spin subsystem through the spin-orbit channel allows to extract information about the orbital subsystem and reduce the entropy of the orbital part.
On the hand result of two incompatible measurements done on the spin subsystem, depends on the fact whether the density matrix of the system is pure or mixed.
In the case of pure states, the spin-orbit coupling works as quantum memory and reduces the uncertainty about the measurement results, whereas, in the case of mixed states, the spin-orbit coupling enhances uncertainty.

\section{Acknowledgment}
We acknowledge financial support from DFG through SFB 762.
	
\newpage

\bibliography{refs}

\end{document}